\documentclass[aps,pre,showpacs,raggedbottom,nobalancelastpage,amssymb,twocolumn,groupedaddress]{revtex4}
\usepackage{graphicx}
\usepackage{amsmath}
\usepackage{amsfonts}
\usepackage{amssymb}
\usepackage{epsfig,libertine}
\usepackage[libertine]{newtxmath}
\usepackage{color}
\usepackage{dsfont, hyperref, xcolor}
\usepackage{comment}
\usepackage{enumerate}

\newcommand{\abs}[1]{\left\vert#1\right\vert}

\newcommand{\Tr}[1]{\text{Tr}\left\{#1\right\}}

\newcommand{\bra}[1]{\langle#1\vert}
\newcommand{\ket}[1]{\vert#1\rangle}
\newcommand\braket[2]{\langle#1|#2\rangle}

\begin{document}

\title{Class of quasiprobability distributions of work and initial quantum coherence}

\author{Gianluca~Francica}
\email{gianluca.francica@gmail.com}
%\address{Via Pozzo - 89844 Nicotera (VV), Italy}

\date{\today}

\begin{abstract}
The work is a concept of fundamental importance in thermodynamics. An open question is how to describe the work fluctuation for quantum coherent processes in the presence of initial quantum coherence in the energy basis. With the aim of giving a unified description, here we introduce and study a class of quasiprobability distributions of work, which give an average work equal to the average energy change of the system and reduce to the two-projective measurement scheme for an initial incoherent state.
Moreover, we characterize the work with the help of fluctuation relations. In particular, by considering the joint distribution of work and initial quantum coherence, we find a fluctuation theorem involving quantum coherence, from which follows a second law of thermodynamics in the case of initial thermal populations. Furthermore, we propose a way to measure the characteristic function of work and we discuss the negativity of the quasiprobability. The effects of coherence are also investigated for a simple system of a qubit.
\end{abstract}

\maketitle
\section{Introduction}
Thermodynamics of physical processes and how they are affected by the quantumness of the nature has received a great attention in the last decades~\cite{campisi11,Vinjanampathy16,book19}. In this context quantum coherence will undoubtedly play a fundamental role, for instance it is strictly related to the irreversible work~\cite{francica19}, it leads a genuine quantum contribution to the ergotropy~\cite{francica20} and makes it possible to create quantum correlations which can be also employed in the work extraction~\cite{francica17}. %Despite recent studies have been performed on the quantum coherence in an operational prospective~\cite{Streltsov17},  its role in thermodynamics processes and how it can be exploited as a quantum resource allowing to obtain processes otherwise classically inhibited certainly remains unclear at all.

Typically, the internal energy of a system changes because work is performed or heat is exchanged with its environment, and the first law of thermodynamics expresses the internal energy change as the sum of work and heat. % (see, e.g., Ref.~\cite{rivas20} for the strong coupling analysis).
Any energy change of the total system composed by the system and the environment, since it is isolated, must be identified with work. Then, in the limit of a weak coupling between system and environment, the internal energy  is equal to the energy  of the system and the heat corresponds to minus the energy change of the environment.
Here we focus on a thermally isolated quantum system such that the energy change of the system is equal to the work performed on the system. Therefore, we consider a typical out-of-equilibrium coherent process performed by control of some external parameters, where the work done follows a statistics which can be also constrained by certain fluctuation theorems for equilibrium initial conditions~\cite{jarzynski97,jarzynski11}.
We recall that the presence of the quantum fluctuation is taken into account by adopting different schemes (see, e.g., Refs.~\cite{roncaglia14,allahverdyan14,solinas15,deffner16,perarnau-llobet17}), and originally a two-projective measurement scheme has been commonly adopted~\cite{talkner07}. %, where a projective energy measurement is performed both at the beginning and at the end of the process.
Anyway, it is well known that in this invasive scheme the first measurement of the energy destroys the initial coherence in the energy basis and a better way to describe the work fluctuation in the presence of initial coherence is by using a quasiprobability distribution.
In particular, we recall that the no-go theorem of Ref.~\cite{perarnau-llobet17} states that there is no scheme having a probability distribution of work, which is linear with respect to the initial state, such that reduces to the two-projective measurement scheme and the average work corresponds to the average energy change. However, the existence of a quasiprobability is admitted. Moreover, any scheme reducing to the two-projective measurement scheme for incoherent states either admits a work quasiprobability or fails to describe protocols exhibiting contextuality~\cite{lostaglio18}.

Thus, in this paper we aim to give a unifying picture by introducing a class of quasiprobability distributions of work. All these quasiprobability distributions give an average work equal to the average energy change of the system and reduce to the two-projective measurement scheme for an initial incoherent state. We characterize this class, also thanks to general relations with the two-projective measurement scheme and fluctuation relations. Furthermore, we explain how the characteristic functions can be measured by looking on the coherence of a detector. In particular, we note that the quasiprobability distributions of Ref.~\cite{allahverdyan14} and Ref.~\cite{solinas15} belong to this class.

\section{Quasiprobability distribution of work}
We consider a quantum coherent process generated through a time-dependent Hamiltonian $H(t)=\sum \epsilon_k(t) \ket{\epsilon_k(t)}\bra{\epsilon_k(t)}$ where $\ket{\epsilon_k(t)}$ is the eigenstate with eigenvalue $\epsilon_k(t)$ at the time $t$. The time evolution operator is $U_{t,0}=\mathcal T e^{-i\int_0^t H(s) ds}$, where $\mathcal T$ is the time order operator and the average work $\langle w\rangle$ done on the system in the time interval $[0,\tau]$ can be identify with the average energy change
\begin{equation}
\langle w\rangle = \Tr{(H^{(H)}(\tau) -H(0) )\rho_0}\,,
\end{equation}
where $\rho_0$ is the initial density matrix and given an operator $A(t)$ we define the Heisenberg time evolved operator $A^{(H)}(t) = U_{t,0}^\dagger A(t) U_{t,0}$.
In our discussion a key role is played by the initial quantum coherence in the energy basis. Given a density matrix, we will say that there is quantum coherence in a certain basis if there are non-zero coherences (i.e. off-diagonal elements of the density matrix) with respect to such basis. This means that a pure state is in a superposition of the states of the basis if there is coherence. Thus, among all the possible states, we can identify the incoherent states with respect to a certain basis as the states having all the coherences equal to zero. Concerning the work, a special case is that of an incoherent initial state $\rho_0$ which is incoherent with respect to the basis of energy eigenstates $\ket{\epsilon_k}=\ket{\epsilon_k(0)}$ and then can be expressed as $\rho_0 = \Delta(\rho_0)$, where we have defined the dephasing map
\begin{equation}
\Delta(\rho_0) = \sum_i \ket{\epsilon_i}\bra{\epsilon_i}\rho_0 \ket{\epsilon_i}\bra{\epsilon_i}\,.
\end{equation}
For an incoherent initial state $\rho_0$ the average work done can be expressed in terms of the two-projective measurement scheme probability distribution
\begin{equation}
 p(w) = \sum_{k,j} \bra{\epsilon_j}\rho_0\ket{\epsilon_j} \abs{\bra{\epsilon'_k}U_{\tau,0}\ket{\epsilon_j}}^2 \delta(w-\epsilon'_k + \epsilon_j)
\end{equation}
as $\langle w\rangle = \int w p(w) dw$, where $\epsilon'_k = \epsilon_k(\tau)$ and $\epsilon_j = \epsilon_j(0)$ (in particular, if the spectrum of $H(0)$ is degenerate, we choose the basis of eigenstates $\ket{\epsilon_j}$ such that the restriction of $\rho_0$ on any eigenspace is diagonal with respect to such basis).
Obviously, this relation does not hold if the initial state $\rho_0$ is not incoherent. % (i.e. there are non-zero coherences).
In general, we define the quasiprobability distribution of work
%\begin{widetext}
%\begin{equation}
%p_q(w) = \sum_{kji} Re\{\bra{\epsilon_i(0)}\rho_0\ket{\epsilon_j(0)} \bra{\epsilon_j(0)}U^\dagger_{\tau,0}\ket{\epsilon_k(\tau)}\bra{\epsilon_k(\tau)}U_{\tau,0}\ket{\epsilon_i(0)}\} \delta(w-\epsilon_k(\tau)+q\epsilon_i(0)+(1-q)\epsilon_j(0))
%\end{equation}
%\end{widetext}
\begin{eqnarray}
\nonumber p_q(w) &=& \sum_{k,j,i} Re\{\bra{\epsilon_i}\rho_0\ket{\epsilon_j} \bra{\epsilon_j}U^\dagger_{\tau,0}\ket{\epsilon'_k}\bra{\epsilon'_k}U_{\tau,0}\ket{\epsilon_i}\}\\
&& \times  \delta(w-\epsilon'_k+q\epsilon_i+(1-q)\epsilon_j)\,,
\end{eqnarray}
where $q$ is a real parameter and of course $p_q(w)=p(w)$ for an incoherent state $\rho_0$.
We note that there is the symmetry relation $p_{1-q}(w)=p_q(w)$. Furthermore, for $q=0,1$ we get the quasiprobability distribution introduced in Ref.~\cite{allahverdyan14} and for $q=1/2$ we get the one of Ref.~\cite{solinas15}. It is easy to show that $\langle w \rangle = \int w p_q(w) dw$ for any $q$ and any initial state $\rho_0$. Then, we will calculate the average of the work $w$ with respect to $p_q(w)$. We have that the second moment is
\begin{equation}
\langle w^2 \rangle = \Tr{(H^{(H)}(\tau)-H(0))^2\rho_0}\,,
\end{equation}
but the higher moments will depend on $q$, for instance the third moment is equal to
%\begin{widetext}
%\begin{equation}
%\langle w^3\rangle = \Tr{\left(\left(H^{(H)}(\tau)-H(0)\right)^3-\frac{1}{2}\left[H^{(H)}(\tau)+ H(0),\left[H^{(H)}(\tau),H(0)\right]\right]+3q(1-q)\left[H(0),\left[H^{(H)}(\tau),H(0)\right]\right]\right)\rho_0}
%\end{equation}
\begin{eqnarray}
\nonumber \langle w^3\rangle &=& \Tr{(H^{(H)}(\tau)-H(0))^3\rho_0}\\
\nonumber && -\frac{1}{2}\Tr{[H^{(H)}(\tau)+H(0),[H^{(H)}(\tau),H(0)]]\rho_0}\\
&&+3q(1-q)\Tr{[H(0),[H^{(H)}(\tau),H(0)]]\rho_0}\,.
\end{eqnarray}
%and in general $\langle w^n \rangle$ will be a polynomial function of $q$ of degree $n-1$.
The characteristic function is defined as $\chi_q(u) = \langle e^{iuw}\rangle$ and reads
%\begin{widetext}
%\begin{equation}
%\chi_q(u) = \frac{1}{2}\bigg(\Tr{e^{-iuqH(0)}\rho_0 e^{-iu(1-q)H(0)} e^{iu H^{(H)}(\tau)}} + \Tr{e^{-iu(1-q)H(0)}\rho_0 e^{-iuqH(0)} e^{iu H^{(H)}(\tau)}}\bigg)
%\end{equation}
%\end{widetext}
\begin{eqnarray}
\nonumber\chi_q(u) &=& \frac{1}{2}\bigg(\Tr{e^{-iuqH(0)}\rho_0 e^{-iu(1-q)H(0)} e^{iu H^{(H)}(\tau)}}\\
 && + \Tr{e^{-iu(1-q)H(0)}\rho_0 e^{-iuqH(0)} e^{iu H^{(H)}(\tau)}}\bigg)\,,
\end{eqnarray}
such that the moments are $\langle w^n\rangle = (-i)^n \partial^n_u\chi_q(0)$.
Conversely, the characteristic function of the two-projective measurement scheme, defined as $\chi(u) = \int e^{iuw}p(w) dw$, can be expressed as $\chi(u) = \Tr{\Delta(\rho_0)e^{-i u H(0)}e^{iu H^{(H)}(\tau)} }$ and does not depend on the initial coherence. We observe that the two characteristic functions are related by the equation
\begin{eqnarray}
\nonumber\chi_q(u) &=& \chi(u)+ \frac{1}{2}\sum_{i\neq j} (e^{-i u (q \epsilon_i+(1-q)\epsilon_j)}+e^{-i u ((1-q) \epsilon_i+q\epsilon_j)}) \\
&& \times \bra{\epsilon_i}\rho_0\ket{\epsilon_j}\bra{\epsilon_j} e^{iu H^{(H)}(\tau)} \ket{\epsilon_i}\,.
\end{eqnarray}
%\begin{widetext}
%\begin{equation}
%\chi_q(u) = \chi(u)+ \frac{1}{2}\sum_{i\neq j} (e^{-i u (q \epsilon_i(0)+(1-q)\epsilon_j(0))}+e^{-i u ((1-q) \epsilon_i(0)+q\epsilon_j(0))}) \bra{\epsilon_i(0)}\rho_0\ket{\epsilon_j(0)}\bra{\epsilon_j(0)} e^{iu H^{(H)}(\tau)} \ket{\epsilon_i(0)}
%\end{equation}
%\end{widetext}
%which differs from $\chi(u)$ because of the coherences in the initial state $\rho_0$ and the coherences produced in the time reversal process with unitary $U^\dagger_{\tau,0}$.
Thus, we note that if $[H^{(H)}(\tau),H(0)]=0$  then $\chi_q(u) = \chi(u)$, the moments are $\langle w^n\rangle = \Tr{(H^{(H)}(\tau)-H(0))^n\Delta(\rho_0)}$ and the initial coherence does not play any role. For instance this is the case of the adiabatic limit. %In particular, in this case we get $p_q(w) = \sum \bra{\epsilon_i(0)}\rho_0\ket{\epsilon_i(0)}\delta(w-\epsilon_i(\tau)+\epsilon_i(0))$, for instance this is the case of the adiabatic limit.

%\section{Fluctuation relations}
In general, we have the fluctuation relation
\begin{equation}\label{fluc rel}
\langle e^{-\beta (w-\Delta F)} \rangle = Re \Tr{e^{\beta q H(0)}\rho_0 e^{-\beta q H(0)} \rho^{-1}_{\beta,0}\rho^{(H)}_{\beta,\tau} }\,,
\end{equation}
where we have defined the initial equilibrium state $\rho_{\beta,0}=e^{-\beta H(0)}/Z_{\beta,0}$, the time-reversed evolved state $\rho^{(H)}_{\beta,\tau} = e^{-\beta H^{(H)}(\tau)}/Z_{\beta,\tau}$, the partition function $Z_{\beta,t}=\Tr{e^{-\beta H(t)}}$ and the free energy change $\Delta F = - \ln(Z_{\beta,\tau}/Z_{\beta,0})/\beta$. Specifically, $\rho^{(H)}_{\beta,\tau}$ is the time evolved final state of the time-reversed process with time evolution operator $U_{\tau,0}^\dagger$ and initial state $\rho_{\beta,\tau}$. We note that the fluctuation relation of Eq.~\eqref{fluc rel} reduces to the one of Ref.~\cite{allahverdyan14} for $q=0,1$.
%It is easy to see that, also in the presence of initial coherence, $\langle e^{-\beta (w-\Delta F)} \rangle \to 1$ in the adiabatic limit if the populations are thermal, i.e. such that $\bra{\epsilon_i(0)}\rho_0\ket{\epsilon_i(0)}=e^{-\beta \epsilon_i(0)}/Z_{\beta,0}$.
%For a density matrix $\rho_0$ with thermal populations, i.e. such that $\bra{\epsilon_i(0)}\rho_0\ket{\epsilon_i(0)}=e^{-\beta \epsilon_i(0)}/Z_{\beta,0}$, we have
%\begin{eqnarray}
%\nonumber \langle e^{-\beta (w-\Delta F)} \rangle &=& 1+ Z_{\beta,0}Re\sum_{i\neq j} e^{\beta q\epsilon_i(0)}e^{\beta (1-q) \epsilon_j(0)}\\
%&& \times  \bra{\epsilon_i(0)}\rho_0\ket{\epsilon_j(0)}\bra{\epsilon_j(0)}\rho^{(H)}_{\beta,\tau}\ket{\epsilon_i(0)}
%\end{eqnarray}
%which differs from one because of the coherences in the initial state $\rho_0$ and the coherences produced in the time reversal process. Thus, it tends to one as the process tends to be adiabatic.

\noindent We proceed our investigation by considering that the initial quantum coherence can be characterized by using the relative entropy of coherence~\cite{Streltsov17}
\begin{equation}
\langle C \rangle = S(\Delta(\rho_0))-S(\rho_0)\,,
\end{equation}
where we have introduced the von Neumann entropy $S(\rho) = - \Tr{\rho \ln \rho}$. By considering the eigenvalues $r_n$ and the eigenstates $\ket{r_n}$ of the initial state $\rho_0$, such that $\rho_0 = \sum r_n \ket{r_n}\bra{r_n}$, we define the probability distribution of coherence
\begin{equation}
p_c(C) = \sum_{i,n} r_n \abs{\braket{\epsilon_i}{r_n}}^2\delta(C+\ln\bra{\epsilon_i}\rho_0\ket{\epsilon_i}-\ln r_n)\,,
\end{equation}
such that $\langle C \rangle=\int C p_c(C) dC$. Therefore, the initial quantum coherence can be thought of as a stochastic variable $C$. % characterized by the probability distribution $p_c(C)$.
The characteristic function is defined as $\chi_c(t) = \langle e^{itC}\rangle$ and reads $\chi_c(t) = \Tr{\rho_0 e^{it\ln\rho_0}e^{-i t \ln \Delta(\rho_0)}}$. Thus, the coherence $C$ is equivalent to the work performed in the process with initial Hamiltonian $-\ln\rho_0$ and time evolved final Hamiltonian $-\ln \Delta(\rho_0)$. Since in this case the free energy change is zero, we get the fluctuation relation $\langle e^{-C}\rangle=1$. Furthermore, we can define the joint quasiprobability distribution
\begin{eqnarray}
\nonumber p_{q,q'}(w,C) &=& \sum_{k,j,i,n} r_n Re\{\braket{\epsilon_i}{r_n}\braket{r_n}{\epsilon_j} \bra{\epsilon_j}U^\dagger_{\tau,0}\ket{\epsilon'_k}\bra{\epsilon'_k}U_{\tau,0}\ket{\epsilon_i}\}\\
\nonumber && \times  \delta(w-\epsilon'_k+q\epsilon_i+(1-q)\epsilon_j)\delta(C+q'\ln\bra{\epsilon_i}\rho_0\ket{\epsilon_i}\\
&&  +(1-q')\ln\bra{\epsilon_j}\rho_0\ket{\epsilon_j}-\ln r_n)\,,
\end{eqnarray}
from which we get the marginal distributions $p_c(C) = \int p_{q,q'}(w,C) dw$ and $p_q(w) = \int p_{q,q'}(w,C) dC$. In particular, in order to be as general as possible, we have introduced another real parameter $q'$.
The quasiprobability distribution $p_{q,q'}(w,C)$ is related to the two-projective measurement scheme by the equation
\begin{equation}
\int e^{-C}p_{q,q'}(w,C)dC = p(w)
\end{equation}
and therefore we have the fluctuation relation
\begin{equation}\label{fluc rel 2}
\langle e^{-\beta(w-\Delta F) -C}\rangle = \Tr{\Delta(\rho_0) \rho^{-1}_{\beta,0}\rho^{(H)}_{\beta,\tau} }\,,
\end{equation}
where the average is calculated with respect to $p_{q,q'}(w,C)$.
It is worth to observe that for an initial state with thermal populations, i.e. such that $\bra{\epsilon_i}\rho_0\ket{\epsilon_i}=e^{-\beta \epsilon_i}/Z_{\beta,0}$, Eq.~\eqref{fluc rel 2} reads
\begin{equation}
\langle e^{-\beta(w-\Delta F) -C}\rangle = 1
\end{equation}
and we have the relation
\begin{equation}\label{2law}
\beta (\langle w \rangle -\Delta F) + \langle C\rangle = S(\rho_\tau || \rho_{\beta,\tau})  \geq 0\,,
\end{equation}
where $\rho_\tau$ is the time evolved state $\rho_\tau=U_{\tau,0} \rho_0 U_{\tau,0}^\dagger$ and $S(\rho || \eta)$ is the quantum relative entropy defined as $S(\rho || \eta)=\Tr{\rho (\ln \rho - \ln \eta)}$. In particular, the inequality of Eq.~\eqref{2law} represents a second law of thermodynamics in the presence of coherence.

\noindent Moreover, by noting that $S(\rho_\tau) = S(\rho_0)$, we have the quantum relative entropy $S(\rho_\tau || \rho_{\beta,\tau}) = - S(\rho_0) - \Tr{\rho_\tau \ln \rho_{\beta,\tau}}$, which can be expressed as the average
\begin{equation}
S(\rho_\tau || \rho_{\beta,\tau}) = \int \sigma p_s(\sigma) d\sigma\,,
\end{equation}
where we have defined the probability distribution
\begin{equation}
p_s(\sigma) = \sum_{k,n} r_n \abs{\bra{\epsilon'_k}U_{\tau,0}\ket{r_n}}^2 \delta(\sigma - \beta \epsilon'_k - \ln Z_{\beta,\tau}-\ln r_n)\,.
\end{equation}
Thus, also the quantum relative entropy can be viewed as a stochastic variable $\sigma$ which satisfies the fluctuation relation $\langle e^{-\sigma} \rangle = 1$. For an initial state with thermal populations, i.e. such that $\bra{\epsilon_i}\rho_0\ket{\epsilon_i}=e^{-\beta \epsilon_i}/Z_{\beta,0}$, we have the relation % with the quasiprobability $p_{q,q'}(w,C)$
\begin{equation}
\int p_{q,q}(w,\sigma-\beta(w-\Delta F))dw = p_s(\sigma)\,.
\end{equation}
This means that for an initial state with thermal populations the variables $\beta (w - \Delta F) + C $  and $\sigma$ have the same statistics if we consider $q'=q$. To understand this, it is enough to consider
\begin{equation}
\beta (\langle w \rangle - \Delta F) + \langle C \rangle = \int (\beta (w - \Delta F) + C) p_{q,q}(w,C) dw dC\,.
\end{equation}
We change variable by defining $\sigma$ such that $C= \sigma - \beta(w-\Delta F) $, then
\begin{eqnarray}
\nonumber \beta (\langle w \rangle - \Delta F) + \langle C \rangle &=& \int \sigma p_{q,q}(w,\sigma - \beta(w-\Delta F)) dw d\sigma \\
 &=& \int \sigma p_s(\sigma) d\sigma = \langle \sigma \rangle
\end{eqnarray}
and in general $\langle f(\beta (w - \Delta F) + C) \rangle = \langle f(\sigma) \rangle$ for any function $f$.

%\section{Measurement and negativity of work distribution}
We note that we can measure $\chi_q(u)$ by proceeding as in Ref.~\cite{solinas15}. We introduce a detector in the initial state $\rho^D_0$ and the time evolution of the total system is generated by $H(t) - \alpha(t) \Lambda \otimes H(t)$ where $\Lambda$ is a detector observable and $\alpha(t) = \delta(t-\tau+0^+)-\delta(t-0^+)$. When the total system is prepared in the initial state $\rho^D_0\otimes \rho_0$, the coherence of the detector state can be expressed as
\begin{equation}
\frac{\bra{\lambda}\rho^D_\tau\ket{\lambda'}}{\bra{\lambda}\rho^D_0\ket{\lambda'}} = \Tr{e^{-i\lambda H(0)}\rho_0 e^{i \lambda' H(0)} e^{i(\lambda -\lambda') H^{(H)}(\tau)}}\,,
\end{equation}
where $\ket{\lambda}$ is the eigenstate of $\Lambda$ with eigenvalue $\lambda$ and $\rho^D_\tau$ is the time evolved detector state. Thus we have
\begin{equation}
\chi_q(u) = \frac{1}{2}\left(\frac{\bra{uq}\rho^D_\tau\ket{u(q-1)}}{\bra{uq}\rho^D_0\ket{u(q-1)}}+ \frac{\bra{u(1-q)}\rho^D_\tau\ket{-uq}}{\bra{u(1-q)}\rho^D_0\ket{-uq}}\right)\,.
\end{equation}

In order to discuss the negativity of the quasiprobability distribution, we define the operator $X_{ijk}$ such that
\begin{equation}
\Tr{X_{ijk}\rho_0} =  Re\{\bra{\epsilon_i}\rho_0\ket{\epsilon_j}\bra{\epsilon_j}U^\dagger_{\tau,0}\ket{\epsilon'_k}\bra{\epsilon'_k}U_{\tau,0}\ket{\epsilon_i}\}\,.
\end{equation}
In general we have
\begin{equation}
-\frac{1}{4}\leq X_{ijk} \leq 1
\end{equation}
(e.g. $-1/4 \leq X_{ijk}$ means that the eigenvalues of $X_{ijk}+1/4$ are non-negative).
To prove this relation, we start by considering $i=j$ such that we have $X_{iik} = \abs{\bra{\epsilon'_k}U_{\tau,0}\ket{\epsilon_i}}^2 \ket{\epsilon_i}\bra{\epsilon_i}$ and, since $0\leq \abs{\bra{\epsilon'_k}U_{\tau,0}\ket{\epsilon_i}}^2\leq 1$, we get $0\leq X_{iik}\leq 1$.
Conversely, for $i\neq j$, we have  $X_{ijk} = \bra{\epsilon_j}U^\dagger_{\tau,0}\ket{\epsilon'_k}\bra{\epsilon'_k}U_{\tau,0}\ket{\epsilon_i} \ket{\epsilon_j}\bra{\epsilon_i}/2 +h.c.$, thus, since $\Tr{X_{ijk}}=0$, the eigenvalues of $X_{ijk}$ are $\pm \sqrt{-\det(X_{ijk})}$, which are $\pm |\bra{\epsilon_j}U^\dagger_{\tau,0}\ket{\epsilon'_k}\bra{\epsilon'_k}U_{\tau,0}\ket{\epsilon_i}|/2$. Since $|\bra{\epsilon_j}U^\dagger_{\tau,0}\ket{\epsilon'_k}\bra{\epsilon'_k}U_{\tau,0}\ket{\epsilon_i}|$ gets its maximum value $1/2$ when $U^\dagger_{\tau,0}\ket{\epsilon'_k} = (\ket{\epsilon_i}+e^{i\phi}\ket{\epsilon_j})/\sqrt{2}$, we get $-1/4 \leq X_{ijk}\leq 1/4$.
%For $i\neq j$ the eigenvalues of $X_{ijk}$ are $\pm |\bra{\epsilon_j}U^\dagger_{\tau,0}\ket{\epsilon'_k}\bra{\epsilon'_k}U_{\tau,0}\ket{\epsilon_i}|/2$, then we have $-1/4 \leq X_{ijk}\leq 1/4$ (e.g. $-1/4 \leq X_{ijk}$ means that the eigenvalues of $X_{ijk}+1/4$ are non-negative). Instead for $i=j$ we have $0 \leq X_{iik}\leq 1$, thus in general we get
%\begin{equation}
%-\frac{1}{4}\leq X_{ijk} \leq 1
%\end{equation}

\section{Physical example}
\begin{figure}
[h!]
\includegraphics[width=0.49\columnwidth]{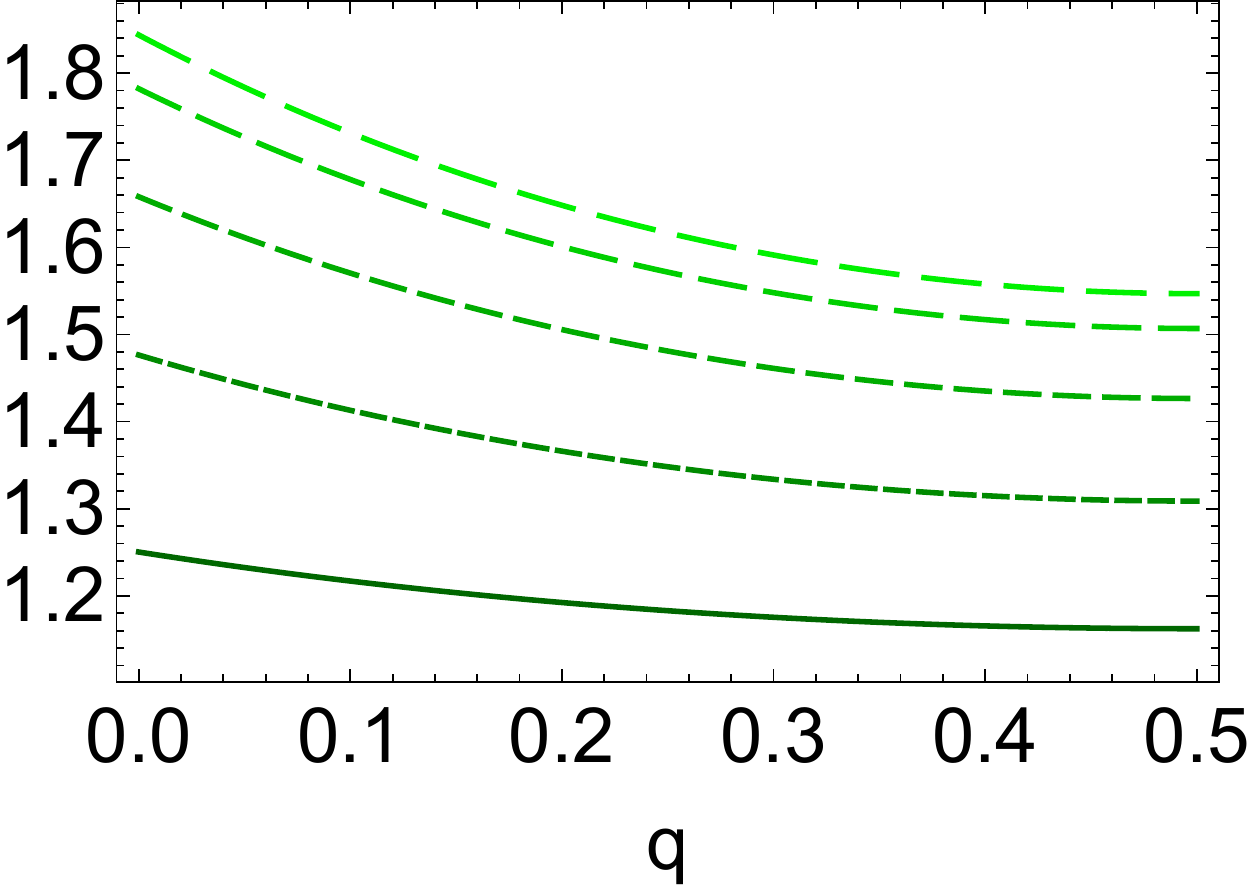}\includegraphics[width=0.49\columnwidth]{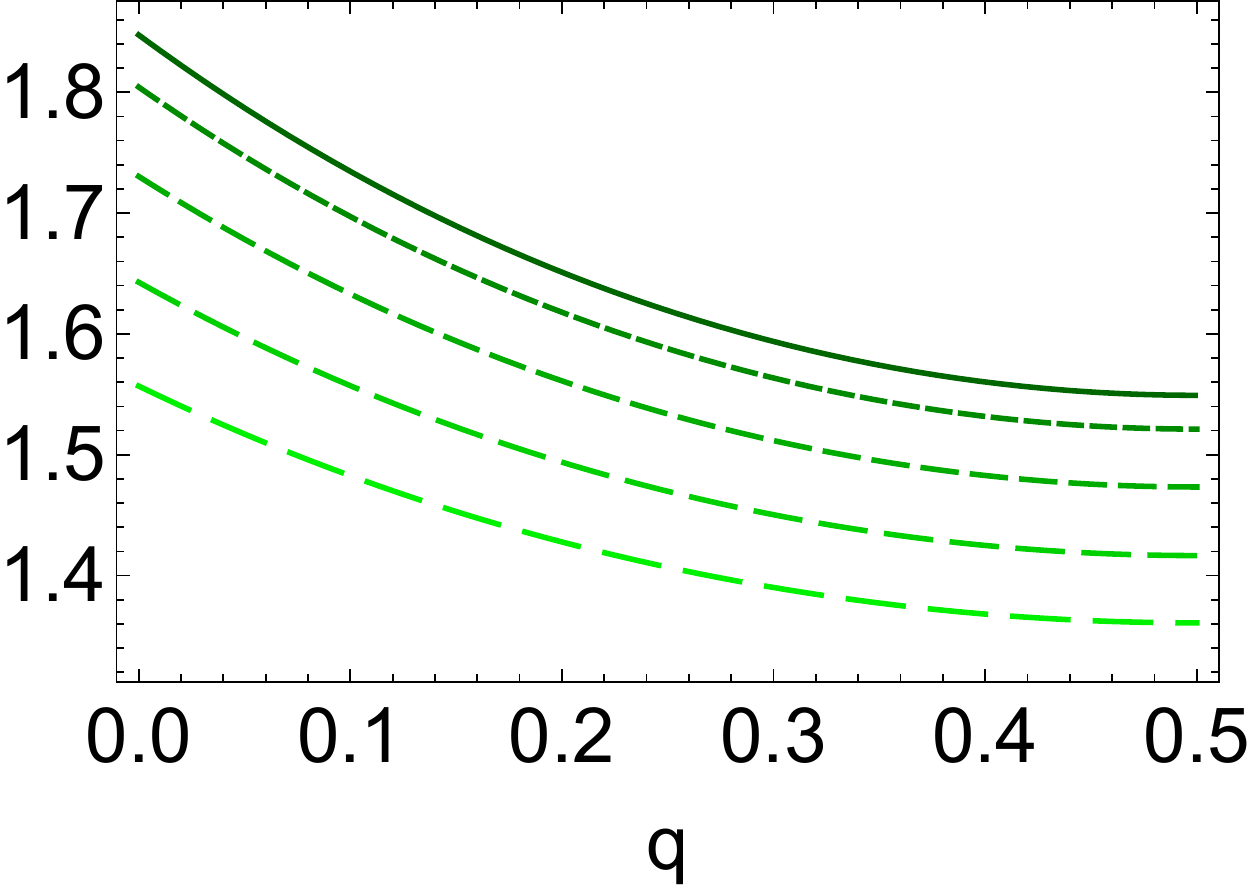}
\caption{ The plots of $\langle e^{-\beta(w-\Delta F)}\rangle$ in function of $q$, for different duration times $\tau$. We consider a maximum coherence $c=\sqrt{p(1-p)}$, i.e. a coherent Gibbs state. We put $\tau\omega_0=0.2,0.4,0.6,0.8,1$ (darker solid to lighter dashed lines, left panel) and $\tau\omega_0=1.2,1.4,1.6,1.8,2$ (darker solid to lighter dashed lines, right panel).
}
\label{fig:plot}
\end{figure}

As a physical example we consider the process experimentally studied in Ref.~\cite{batalhao15}, which is a qubit with Hamiltonian $H(t) = \omega(t) (\sigma^x \cos \varphi(t)+\sigma^y \sin \varphi(t))$, where $\varphi(t) = \pi t/(2\tau)$, $\omega(t) = \omega_0 (1-t/\tau)+\omega_\tau t/\tau$ and $\sigma^x$, $\sigma^y$ and $\sigma^z$ are the Pauli matrices. For studying the effect of the initial coherence we will take the initial density matrix $\rho_0 = I/2+ (2p-1) \sigma^x/2 + c \sigma^z$. The characteristic function reads
\begin{equation}
\chi_q(u) = \chi(u) + c \left( \cos(2 u q \omega_0) a(u) -\sin(2 u q \omega_0) b(u) \right)\,,
\end{equation}
where we have defined the complex functions
\begin{eqnarray}
a(u) &=& \frac{1}{2}\Tr{\sigma^z \{e^{-iu \omega_0 \sigma^x},e^{iu \omega_\tau U^\dagger_{\tau,0}\sigma^y U_{\tau,0}}\}}\,,\\
b(u) &=& \frac{1}{2}\Tr{\sigma^y [e^{-iu \omega_0 \sigma^x},e^{iu \omega_\tau U^\dagger_{\tau,0}\sigma^y U_{\tau,0}}]}\,.
\end{eqnarray}
%and the characteristic function
%\begin{equation}
%\chi(u) = \Tr{\left(\frac{I}{2}+ \frac{2p-1}{2} \sigma^x \right)e^{-iu \omega_0 \sigma^x}e^{iu \omega_\tau U^\dagger_{\tau,0}\sigma^y U_{\tau,0}}}
%\end{equation}
%By considering thermal populations, i.e. $p=e^{-\beta \omega_0}/Z_{\beta,0}$, we get $\chi(i\beta) = e^{-\beta \Delta F}$ thus the fluctuation relation reads
%\begin{equation}
%\langle e^{-\beta w}\rangle = e^{-\beta\Delta F}+ c \left( \cosh(2 \beta q \omega_0) a(i\beta) -i\sinh(2 \beta q \omega_0) b(i\beta) \right)
%\end{equation}
We investigate the fluctuation relation of Eq.~\eqref{fluc rel} by considering a state with thermal populations $p=e^{-\beta \omega_0}/Z_{\beta,0}$ and the values $\omega_\tau=2\omega_0$ and $\beta \omega_0 =1$. Then, we note that $\langle e^{-\beta(w-\Delta F)}\rangle$ tends to one for a sudden quench $\tau \omega_0 \to 0$, increases with $\tau$ until $\tau \omega_0 \approx 1$, then decreases such that as $\tau$ tends to infinity we get $\langle e^{-\beta(w-\Delta F)}\rangle \to 1$ (see Fig.~\ref{fig:plot}).
%For $q\leq 1/2$, $\langle e^{-\beta(w-\Delta F)}\rangle$ is a decreasing function of $q$, then reaches its minimum (is closer to one) at $q=1/2$.
This behavior can be understood by considering that the time evolution acts as a spin rotation, such that $U^\dagger_{\tau,0}\sigma^y U_{\tau,0} = \hat n \cdot \vec \sigma$ with $\hat n$ unit vector. Therefore, the functions $a(u)$ and $b(u)$ read
\begin{eqnarray}
a(u) &=& 2i n_z \cos(u\omega_0) \sin(u\omega_\tau)\,, \\
b(u) &=& -2i n_z \sin(u\omega_0) \sin(u\omega_\tau)
\end{eqnarray}
and we get the characteristic function
\begin{equation}
\chi_q(u) = \chi(u) + 2 i c n_z\cos(2u(q-1/2)\omega_0)\sin(u\omega_\tau)\,.
\end{equation}
For a sudden quench we have $U_{\tau,0}=I$ such that $\hat n \cdot \vec \sigma = \sigma^y$. Conversely, in the adiabatic limit we get $\hat n \cdot \vec \sigma = \sigma^x$. Then, in both cases $n_z=0$ and thus $\chi_q(u) =\chi(u)$ and $\langle e^{-\beta(w-\Delta F)}\rangle = 1$ for thermal populations. Thus, in order to get a contribution of the initial quantum coherence we need to rotate the spin outside the $xy$-plane generating the component $n_z$ at the end of the time evolution. Furthermore, it is evident that, for thermal populations, $\langle e^{-\beta(w-\Delta F)}\rangle$ is closer to one for $q=1/2$.

\section{Conclusions}
As an energy measurement destroys the initial quantum coherence in the energy basis, it is an open question in quantum thermodynamics that how to describe the work fluctuation for quantum coherent processes beyond two-projective measurement scheme. In summary, here we have approached this problem by defining and investigating a class of quasiprobability distributions of work, giving an average work equal to the average energy change of the system and reducing to the two-projective measurement scheme for an initial incoherent state.
In particular, we have found two different fluctuation relations characterizing the work. The first is the analogous to the one derived in Ref.~\cite{allahverdyan14}. The second is obtained by considering the joint distribution of work and initial quantum coherence and thus involves quantum coherence. Furthermore, for an initial state with thermal populations this joint distribution is intimately related to the probability distribution of quantum relative entropy, from which follows a second law of thermodynamics. % (by applying the Jensen's inequality).
We have also proposed a way to measure the characteristic function by using a detector and discussed the negativity of the quasiprobability.
In conclusion, our work provides new results in the field of quantum thermodynamics and we hope that it will inspire further investigations and applications of quantum coherence in such field.

\end{document}